\documentclass[prb,twocolumn,showpacs]{revtex4}
\usepackage{graphicx}
\usepackage{amsmath}
\usepackage{amssymb}
\usepackage{bbold}

\newcommand{\eps}{\varepsilon}

\newcommand{\ra}{\rangle}
\newcommand{\la}{\langle}
\DeclareMathAlphabet{\bi}{OML}{cmm}{b}{it}

\begin{document}
\title{Magnetotransport properties of a magnetically modulated 
two-dimensional electron gas with the spin-orbit interaction}
\author{SK Firoz Islam and Tarun Kanti Ghosh}
\affiliation{Department of Physics, Indian Institute of Technology-Kanpur,
Kanpur-208 016, India}

\begin{abstract}
We study the electrical transport properties of a two-dimensional 
electron gas with the Rashba spin-orbit interaction in presence of 
a constant perpendicular magnetic field $(B_0 \hat z)$ which is 
weakly modulated by ${\bf B_1} =  B_1 \cos (q x) \hat z$, where 
$B_1 \ll B_0 $ and $q = 2 \pi/a $ with $a$ is the modulation period. 
We obtain the analytical expressions of the diffusive conductivities 
for spin-up and spin-down electrons. The conductivities for spin-up 
and spin-down electrons oscillate with different frequencies and
produce beating patterns in the amplitude of the Weiss and 
Shubnikov-de Haas oscillations. 
We show that the Rashba strength can be determined by analyzing the
beating pattern in the Weiss oscillation.
We find a simple equation which determines the Rashba spin-orbit 
interaction strength if the number of Weiss oscillations between any 
two successive nodes is known from the experiment. We compare our results 
with the electrically modulated 2DEG with the Rashba interaction.
For completeness, we also study the beating pattern formation 
in the collisional and the Hall conductivities.
 
\end{abstract}


\date{\today}

\maketitle
\section{INTRODUCTION}
The magnetotransport properties of a two-dimensional electron gas (2DEG) 
in presence of a weakly modulated one-dimensional (1D) periodic electric 
potential has been studied in great details experimentally and theoretically 
for a long time \cite{Weiss,Gerh,poulo,kotha,vasilo}.
In the absence of the modulation and the spin-orbit interaction (SOI), 
the magnetoconductivity oscillations due to the charged impurities are 
commonly known as the Shubnikov-de Hass (SdH) oscillations.
In the presence of the modulation and at low magnetic fields, it has been 
observed that the magnetoresistivity tensor oscillates with the inverse of 
the magnetic field. These oscillations are completely different in periodicity 
and temperature dependence from the SdH oscillations observed at higher magnetic 
field. These oscillations are commonly known as the Weiss oscillations. 
This is due to the effect of the commensurability between the two length 
scales in the system:
the cyclotron diameter at the Fermi energy and the modulation period  $a$.
The similarity and differences between the Weiss and the SdH oscillations 
are as follows: i) both the oscillations are periodic in $1/B_0$ ii)
the period of the Weiss oscillation varies with the electron density ($n_e$) 
as $ \sqrt{n_e} $, whereas that of the SdH ones as $n_e$;
iii) the amplitude of the Weiss oscillations less dependent on temperature than
that of the SdH oscillations; iv) the Weiss oscillations are visible at weak
magnetic field ($B_0 < 0.6$ T) and the SdH oscillations modulated by the Weiss
oscillations are visible at higher fields. 

In conventional 2DEG systems, magnetotransport properties in the 
presence of 1D electric and magnetic modulations are continuing 
to be an active research field. The magnetotransport properties of 
a 2DEG in presence of a weakly modulated magnetic field has been 
studied theoretically \cite{peeters,super,zhong,matuli,russia,xia,papp}. 
It was theoretically observed that the magnetoconductivity oscillates with 
inverse of the magnetic field and amplitude of the oscillation is much higher 
compare to the case of electrical modulation. 
The magnetothermodynamical properties of electrically or magnetically 
modulated 2DEG has also been studied where the Weiss-type oscillation is 
also shown \cite{vasilo,stewart,tong}.
Later, the magnetoresistance oscillation in a magnetically modulated 2DEG system
was observed experimentally \cite{mag_exp1,mag_exp2,mag_exp3}.
In these experiments, a 1D spatially varying magnetic modulation was achieved by
placing micro-patterned ferromagnet or superconductor on the surface of a 2DEG system.

An internally generated crystal field induces the 
SOI which is known as the Rashba SOI. 
The Rashba interaction strength can also be controlled by a 
strong external electric field acting normal to the 2DEG plane. 
Using the Rashba interaction it was possible to explain 
many experimentally observed features like the combined resonances 
\cite{stein,chang} and the beating patterns in the SdH oscillations 
\cite{miller}.
The Rashba interaction is responsible for many other novel effects
like the spin-FET \cite{datta}, spin-galvanic effect \cite{galvanic}
and the spin Hall effect \cite{perel,she}. 
The Rashba SOI in the 2DEG influences  
various properties such as transport \cite{vasilo_trans,sigi}, 
magnetotransport \cite{vasi,yang,novo,firoz}, magnetization \cite{zhang} 
etc. In Refs. \cite{wang,spin,perov}, the effect of the Rashba interaction 
on magnetotransport properties of electrically modulated 2DEG have been
studied theoretically.

Generally, beating pattern analysis in the SdH oscillation \cite{miller} and 
weak anti-localization method \cite{antiloca} are used to extract the strength of 
the Rashba SOI in the 2DEG system. However, in the presence of $B_0 \neq 0$, 
Zeeman splitting is also accompanied with the Rashba spin splitting. To avoid this problem
optical measurement \cite{optical} has also been proposed.  The Weiss oscillation is
due to the modulation induced Landau levels broadening which
oscillates with inverse magnetic field. Moreover, this oscillation appears
at low magnetic field where the Rashba SOI dominates over the Zeeman splitting.
The Weiss oscillation is less influenced by the Zeeman term specially in the periodicity.
To determine the Rashba strength, beating pattern analysis in the 
Weiss oscillation is more reliable than the SdH oscillation.

In this work, our primary goal is to study analytically the 
Weiss oscillations of the 2DEG with the Rashba interaction
and to determine the strength of the Rashba SOI by analyzing 
the beating pattern of the Weiss oscillations. 
For completeness, we also study other transport coefficients such as
the collisional and the Hall conductivities of the spin-up and spin-down 
electrons in details.

This paper is organized as follows. In section II, we summarize the 
energy eigenvalues, the corresponding eigenfunction and density of
states of the 2DEG with the Rashba interaction in presence of a 
uniform magnetic field.
Also, the first-order energy corrections due to the magnetic 
perturbations and group velocities for spin-up and spin-down 
electrons are evaluated. We calculate and discuss the diffusive, 
collisional and the Hall conductivities by using the semi-classical 
Kubo formula in section III. The summary of our work is presented
in section IV.

\section{ENERGY EIGENVALUE, EIGENFUNCTION AND DENSITY OF STATES IN 
PRESENCE OF the RASHBA SOI}
We consider a 2DEG in the $x$-$y$ plane subjected to a magnetic field 
$ {\bf B} = [B_0 + B_1 \cos(q x)] \hat{z}$, where $ q = 2\pi/a $ and
$ a $ is the modulation period. Also, we consider strength of the magnetic 
modulation is very weak i.e., $B_1 \ll B_0 $. In the Landau gauge the corresponding 
vector potential ${\bf A} $ is $ {\bf A} =  [B_0 x + (B_1/q) \sin(q x)] \hat y $.

The Hamiltonian of an electron with charge $-e$ in presence of the 
perpendicular magnetic field ${\bf B} $ is
\begin{equation}
H = \frac{({\bf p} + e {\bf A})^2}{2m^{\ast }} \mathbb{1} + 
\frac{\alpha}{\hbar }\left[ {\mbox{\boldmath $\sigma$} } \times 
({\bf p}+e{\bf A})\right]_z + \frac{g}{2} \mu _{_B} {\bf B} \cdot 
{\mbox{\boldmath $\sigma$}}, \label{Ham}
\end{equation}
where ${\bf p}$ is the 2D momentum operator, $m^{\ast }$ is the effective mass 
of the electron, $g$ is the Lande $g$-factor, $\mu_{_B}$ is the Bohr magneton,
$\mathbb{1} $ is the identity matrix, 
${\mbox{\boldmath $\sigma$} }=(\sigma_x,\sigma_y,\sigma_z)$ are the Pauli 
spin matrices, and $\alpha $ is the strength of the Rashba interaction. 

Expanding the above Hamiltonian and re-writing as a sum of the
various Hamiltonians:
$ H = H_0 + H_1 + H_2 + H_3  + H_4$, where
\begin{eqnarray}
H_0 & = & \frac{1}{2m^{\ast}}[p_x^2 + (p_y + e B_0 x)^2] \mathbb{1} 
+ \frac{\alpha}{\hbar} [\sigma_x(p_y + e B_0 x) \nonumber \\ 
& - & \sigma_y p_x] + \frac{1}{2} g \mu_{_B} B_0 \sigma_z,
\end{eqnarray}
\begin{eqnarray}
H_1 & = & \frac{V_{_B}}{\hbar q} (p_y + e B_1 x) \sin(qx) \mathbb{1}, \\
H_2 & = & {V_{_B}} \frac{p_{\alpha}}{2\pi} \sigma_x \sin(qx),\\
H_3 & = & V_{_B}g^*\cos(qx) \sigma_z,\\
H_4 & = & \frac{V_{_B}^2}{(4\pi)^2 \epsilon_a}\{1-\cos(2qx)\} \mathbb{1}.
\end{eqnarray}
Here, $V_{_B} = \hbar \omega_1 = \hbar e B_1/m^{\ast} $ is the 
strength of the effective magnetic potential determined by the 
amplitude $B_1$ of the magnetic modulation and $\omega_1 \ll \omega_0 $ 
with $\omega_0 = e B_0/m^{\ast}$ is the cyclotron frequency due to the
constant magnetic field $B_0$. 
The dimensionless parameter $ p_{_\alpha} = a k _{_\alpha} $ with 
$k_{_\alpha} = \alpha m^{\ast}/\hbar^2$ and  $g^*=gm^*/(4m_0)$. 
Also, $ \epsilon_a = \hbar^2/(m^{\ast} a^2) $ is a characteristic energy 
scale introduced by the modulation period $a$.   
Here, the term $H_0 $ is the Hamiltonian for the electron with
the Rashba interaction in presence of the constant magnetic field
$B_0 \hat z$ including the Zeeman energy.
The terms $H_1$ and $H_4$ are due to effect of the magnetic modulation on 
the kinetic energy term. The term $H_2 $ is the interaction between
the spin-orbit and the magnetic modulation whereas the term $H_3$ is the 
interaction between the Zeeman energy and the magnetic modulation. 

The Hamiltonian $H_0$ can be solved analytically and can be treated 
as a unperturbed Hamiltonian. The other four Hamiltonians can be 
treated as a small perturbations since $ V_{_B} \ll \hbar \omega_0$. 
The eigenfunctions of the unperturbed Hamiltonian $(H_0)$ can be 
used to find the energy correction due to the small perturbations 
$H_1, H_2, H_3 $ and $H_4$.

Here, we shall briefly summarize the results of the Ref. \cite{vasi} 
where the analytical solutions of the Hamiltonian $H_0 $ have been derived.
Using the Landau wave functions without the Rashba interaction
as the basis, one can write new wave function as
\begin{equation}
\Psi _{k_y}({\bf r}) = \frac{e^{ik_y y}}{\sqrt{L_y}} 
\sum_{n=0}^\infty \phi_n(x + x_0)\left(\begin{array}{c}
C_n^+ \\
C_n^-
\end{array}
\right). 
\label{wav}
\end{equation}
Here, $\phi_n(x) = (1/\sqrt{\sqrt{\pi }2^n n!l_0}) 
e^{-x^2/2l_0^2} H_n(x/l_0) $ is the normalized harmonic 
oscillator wave function with $n$ is the Landau level index,
$L_y $ is the width of the sample in the $y$ direction, 
$l_0 = \sqrt{\hbar /(e B_0)}$ is the magnetic length, 
and the cyclotron orbit is centered at $-x_0$ with $x_0 =  k_y l_0^2 $.

Using these wave functions and Eq.(1), the eigenvalue problem
$H_0 \Psi = E \Psi $ leads to an infinite number of equations that
can be solved exactly after decomposing it into independent systems
of one or two  equations \cite{vasi}. The resulting eigenstates are
labeled by a new quantum number $s$ instead of $n$.
For $s=0$, there is only one level, the same as the lowest Landau
level without SOI, with energy
$ E_0^+ = E_0 = (\hbar \omega_0 - g \mu _{_B} B_0)/2 $
and the corresponding wave function is
\begin{equation}
\Psi_{0,k_y}^+({\bf r}) = \frac{e^{i k_y y}}{\sqrt{L_y}} \phi_0(x + x_0)
\left(
\begin{array}{c}
0 \\
1
\end{array}
\right).
\end{equation}
For $s=1,2,3....$ there are two branches of energy levels, denoted by $+$ 
and $-$ with energies
\begin{equation}
E_s^{\pm} = s\hbar\omega_{0}{\pm} \sqrt{E_0^2 + s E_{\alpha} \hbar \omega_0},
\end{equation}
where $ E_{\alpha} = 2 m^{\ast} \alpha^2/\hbar^2$.
These energy levels are degenerate in $k_y$. Therefore, the group velocities,
$v_x$ and $v_y$, are zero.

The corresponding wave function for $+$ branch is
\begin{equation}
\Psi _{s,k_y}^+({\bf r}) = \frac{e^{i k_y y}}{\sqrt{L_y A_s }}\left(
\begin{array}{r}
D_s \phi_{s-1}(x + x_0)
\\
\phi_s (x + x_0)
\end{array}
\right) \text{,}
\end{equation}
and for $-$ branch is
\begin{equation}
\Psi_{s,k_y}^-({\bf r})=\frac{e^{i k_y y}}{\sqrt{L_y A_s }}\left(
\begin{array}{r}
\phi_{s-1}(x + x_0)
\\
- D_s \phi_s(x + x_0)
\end{array}
\right) \text{,}
\end{equation}\\
where $A_s = 1 + D_s^2 $ and
$ D_s = \sqrt{s E_{\alpha} \hbar \omega_0}/
[E_0 + \sqrt{E_0^2 + s E_{\alpha} \hbar \omega_0}]$.

The Landau level quantum numbers $ s^{\pm}$ at the Fermi energy are
determined from the equation
$ E_{_F} = s \hbar \omega_0 \pm \sqrt{E_0^2 +  s E_{\alpha} \hbar \omega_0} $.
These quantum numbers $s^{\pm}$ are given by
$ s^{\pm} \simeq k_{_F}^2 l_0^2/2 \mp k_{\alpha} k_{_F} l_0^2$,
where $k_{_F} = \sqrt{2 \pi n_e} $ is the Fermi wave vector and
$n_e$ is the electron density.

The approximate density of states (DOS) of the 2DEG with 
the Rashba SOI in the presence of constant perpendicular magnetic field 
is given \cite{firoz} by
\begin{eqnarray} \label{dos}
D^{\pm}(E) & \approx & \frac{m^*}{2\pi \hbar^2}
\Big[1+2\exp{\Big\{-2\Big(\frac{\pi\Gamma_0}{\hbar\omega_0}\Big)^2\Big\}}
\nonumber \\ & \times &
\cos{\Big\{\frac{2\pi}{\hbar\omega_0}\Big(E+\frac{E_{\alpha}}{2}\mp
\sqrt{E_0^2+E_{\alpha}E}\Big)\Big\}}\Big],
\end{eqnarray}
where $ \Gamma_0 $ is the Landau level broadening.
The detail derivation of the DOS is given in the Appendix A. 
We shall use the above DOS to obtain the analytic expressions of the
magnetotransport coefficients in the presence of the modulations.

Using the perturbation theory, we calculate the first-order energy 
correction for $+$ branch as well as $-$ branch.
It is to be noted that the correction due to the Hamiltonian $H_4$ is of
the order of $\omega_1^2 $ which is much smaller than $ \omega_0$. Therefore,
we neglect the contribution of $H_4$ to the total energy correction.
 
The total energy for $+$ branch and $-$ branch is given by
\begin{eqnarray}
E_{s,k_y}^{\pm}  & = &  s \hbar \omega_0  \pm \sqrt{E_0^2 + 
s E_{\alpha} \hbar \omega_0} + \cos(qx_0) F_s^{\pm}(u), 
\end{eqnarray}
where
\begin{eqnarray}
F_s^{\pm}(u) & = & \frac{V_B}{A_s}\Big[ e^{-u/2}g^{*}
\Big\{\pm D_s^2 L_{s-1/2 \mp 1/2}(u) \nonumber \\&\mp& L_{s -1/2 \pm 1/2}(u)\Big\}
 + D_s^2 G_{s-1/2 \mp 1/2}(u) \nonumber\\&+& G_{s -1/2 \pm 1/2}(u) 
\pm  \frac{Z}{\sqrt{2s}} e^{-u/2} L_{s-1}^1(u)\Big].
\end{eqnarray}
Here, $u=q^2l_{0}^2/2$,  $s=0,1,2...$ for $+$ branch and $s=1,2,3...$ for 
$-$ branch. Also,
$ G_s(u) = e^{-u/2}[L_{s}(u)/2 + L_{s-1}^1(u)] $ and  
$Z=2k_{\alpha} D_{s} l_{0}$. 
The upper and lower signs correspond to the $+$ and $-$ branches, respectively.
The width of the broadened levels of the two branches is 
$ |\Delta_s^{\pm} |$ and it can be written as $ \Delta_s^{\pm} = 2 F_s^{\pm}(u) $.

For various plots, we use the following parameters: 
$\alpha$ in units of $\alpha_{0}= 1.0 \times 10^{-11} $ eV-m, modulation
strength $V_{_B}=0.05$ meV for $B_1=0.02$ T, electron density 
$n_e =10^{16}$/m${}^2$, electron effective mass 
$m^{\ast}=0.05m_0$ with $m_0$ is the free electron mass, 
mobility $ \mu=100$ m${}^2$/V-s, $g=2$,
modulation period $a = 800 $ $\AA$ and temperature $ T = 1.5 $ K.
For these parameters, $p_{_\alpha}=1.05 $ for $\alpha = 2\alpha_0$, 
$p_{_F}=  a k_{_F} = 20 $, $ E_{_F} = 47.5$ meV 
and $\epsilon_a =0.236 $ meV. 

In Fig. 1, we plot the dimensionless bandwidth $\Delta^{\pm}/V_B$ at the Fermi
energy as a function of the dimensionless inverse magnetic field 
$\lambda = B_a/B_0$ for two different
values of $\alpha$. Here, $B_a = \hbar/(ea^2) = 0.102 $ T is the 
characteristic magnetic field introduced by the modulation period $a$.
We plot $\Delta^{\pm}$ instead of $ | \Delta^{\pm}| $ to show the oscillations
more clearly.
\begin{figure}[t]
\begin{center}\leavevmode
\includegraphics[width=97mm]{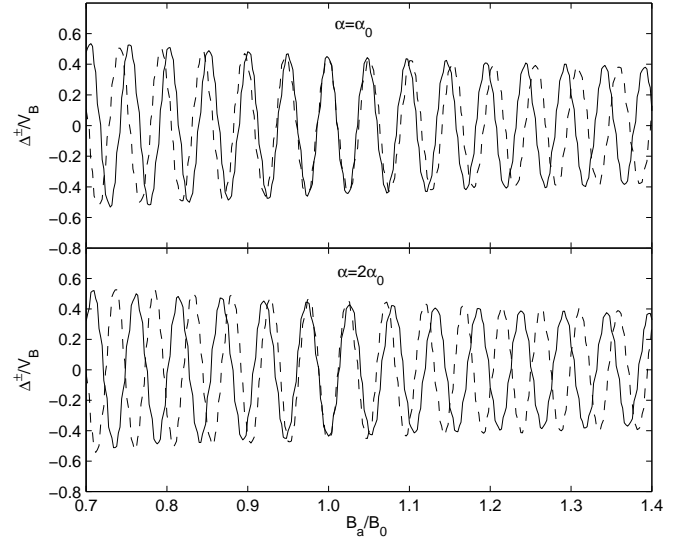}
\caption{Plots of the bandwidth $ \Delta^{\pm}/V_{B}$ at the Fermi energy versus
dimensionless inverse magnetic field $B_a/B_0$ for two different values
of $\alpha$. Here, solid and dashed lines correspond to the upper branch and 
lower branch, respectively.}
\label{Fig1}
\end{center}
\end{figure}

The diagonal matrix elements of the velocity operator do not vanish due
to the $k_y$ dependence of the energy levels.
It can be calculated by using 
$v_y^{\pm} = (1/\hbar)\partial E_{s,k_y}^{\pm}/\partial k_y $, and their 
values are given by 
\begin{equation} \label{velo}
v_y^{\pm} = - \frac{q l_0^2}{\hbar} \sin(qx_0) F_s^{\pm}(u).
\end{equation}
These non-zero values give rise to the finite diffusive conductivity
whereas in the absence of the magnetic modulation the diffusive
conductivity vanishes whether the Rashba interaction is present or not.
These velocities for spin up and spin down branches do oscillate 
with slightly different frequencies as a function of the magnetic field $B_0$.

\section{Magnetotransport Coefficients}
In the presence of the weak modulation, there are two 
contributions to the transport properties:  diffusive 
and collisional contributions. The former is due to 
finite drift velocity gained by the electrons in the presence of 
the modulation. The latter is the contribution from the 
hopping of the localized states due to scattering by impurities.
The diffusive conductivity decreases with increasing impurity 
scattering whereas the collisional conductivity increases with 
the increase of the impurity scattering. The collisional scattering 
is dominant at low temperature.

We follow the formulation of Ref.\cite{vilet} to calculate 
the magnetotransport coefficients in the presence of the modulation. 
In the linear response regime and weak scattering potentials, 
the conductivity tensor, in the one-electron approximation, 
is a sum of a diagonal and non-diagonal terms 
$i.e.$ $\sigma_{\mu \nu} =  \sigma_{\mu \nu}^{\rm d}
+ \sigma_{\mu \nu}^{\rm nd} $, where $ \mu, \nu = x,y$. 
Generally, the diagonal conductivity contains both diffusion 
and collisional contributions: 
$\sigma_{\mu \nu}^{\rm d} = \sigma_{\mu \nu}^{\rm dif} + 
\sigma_{\mu \nu}^{\rm col}$, whereas the Hall conductance is 
obtained from the non-diagonal conductivity 
$\sigma_{\mu \nu}^{\rm nd} $. 
In our problem, $ \sigma_{xx} = \sigma_{xx}^{\rm col} $ because 
$\sigma_{xx}^{\rm dif} = 0 $
and $ \sigma_{yy} = \sigma_{yy}^{\rm col} + \sigma_{yy}^{\rm dif} $. 
The Weiss oscillation is observed in the diagonal component of 
the magnetoresistance $ \rho_{xx} $ which is inverse of the 
conductivity tensor:
$ \rho_{xx} = \sigma_{yy}/(\sigma_{xx} \sigma_{yy} - 
\sigma_{xy} \sigma_{yx})$. 
The effect of the modulation in the transport coefficients, 
$ \sigma_{xx}^{\rm col}$ and $ \sigma_{xy} $, 
is very weak.
Therefore, the change in the conductivities due to the modulation 
[$ \Delta \sigma_{\mu \nu} = \sigma_{\mu \nu}(V_B) - \sigma_{\mu \nu}(0) $]
as a function of the magnetic field will be shown.

\subsection{Diffusive conductivity}
The standard semi-classical expression for the diffusive conductivity 
is given as \cite{vilet}
\begin{equation}
\sigma_{yy}^{\rm dif} = \frac{\beta e^2}{S_0}\sum_{\xi} f_{\xi}(1-f_{\xi})
\tau(E_{\xi})(v_y^{\xi})^2 ,
\end{equation}
where $S_0$ is the area of the system, $\tau$ is the electron relaxation time,
$ f(E)$ is the Fermi-Dirac distribution function, and $\beta=(1/k_{_B} T)$ is 
the inverse of the thermal energy. Also, $ \xi $ is the set of the quantum numbers:
$ \xi: \{s, \sigma, k_y \}$ with $\sigma = \pm $ and 
$ v_y^{\xi} = \la \xi | \hat v_y | \xi \ra $ is the diagonal matrix elements of the 
velocity operator $\hat v_y$.

The non-zero expectation value of the $y$-component of the 
velocity operator is reflected in the finite value diffusive conductivity
whether the SOI is considered or not. 
In the presence of the Rashba interaction, the two energy branches, 
$+$ branch and $-$ branch, contribute to the total conductivity. 
The diffusive conductivity for $\pm$ branches  
is given as  
\begin{equation} \label{conduc}
\sigma_{yy}^{{\rm dif}(\pm)} = \frac{\beta e^2\tau}{S_0} 
\sum_s\int_0^{L_x/l_0^2} dk_y
f(E_s^{\pm})[1-f(E_s^{\pm})](v_y^{\pm})^2.
\end{equation}
In the limit of weak magnetic field $B_0$, many Landau levels are filled
($ s\gg 1$), we use the following approximations: 
$e^{-u/2} L_s(u) \simeq (\pi^2 su)^{-1/4} \cos(2\sqrt{su} - \pi/4) $ and
$\sum_s \rightarrow \int_0^{\infty} dE/(\hbar \omega_0) $.
With the use of the above approximations and following Ref. \cite{vasilo}, 
Eq. (\ref{conduc}) 
reduced to an analytical form which is purely Weiss contribution, as 

\begin{eqnarray} \label{diff_con}
\sigma_{\rm Weiss}^{\pm} & \backsimeq & \frac{e^2}{h} 
A_{_B} C_0^{\pm} \lambda  
\Big[ 1 + \Big( \frac{W_0^{\pm}}{C_{0}^{\pm}} \Big) 
H\Big(\frac{T}{T_a}\Big) \nonumber \\ 
& \times & \sin \big( 2\pi\Omega^{\pm} \lambda  
- \Phi_0^{\pm} \big) \Big],
\end{eqnarray}
where $\Omega^{\pm} = 2(p_{_F} \mp p_{_\alpha})$ are  
the frequencies of the conductivity oscillations of spin-up and spin-down
electrons, $A_{_B}=V_{_B}^2\tau/(\hbar\eps_a)$,
 $T_a = E_F/(2\pi^2 k_{_B} p_{_F} \lambda)$ 
is the characteristic temperature.
and $\Phi_0^{\pm}=\delta_0^{\pm} + 4\pi/(p_{_F}\mp p_{\alpha})$
are the phase factors. 
Other parameters are $ H(x) = x/\sinh(x)$, 
\begin{eqnarray}
C_0^{\pm} &=& \frac{1}{4(p_{_F}\mp p_{_\alpha})} + 
\frac{(p_{_F}\mp p_{\alpha})^3}{16 \pi^4}\sin^2 
\Big(\frac{2\pi}{p_{_F}\mp p_{_\alpha}}\Big),\nonumber
\end{eqnarray}
\begin{eqnarray}
C_1^{\pm} & = & \frac{1}{4(p_{_F}\mp p_{_\alpha})}-
\frac{(p_{_F}\mp p_{\alpha})^3}{16 \pi^4} 
\sin^2\Big(\frac{2\pi}{p_{_F}\mp p_{_\alpha}}\Big),\nonumber
\end{eqnarray}
and
\begin{eqnarray}
C_2^{\pm} & = & \frac{(p_{_F}\mp p_{\alpha})}{4\pi^2}
\sin\Big(\frac{2\pi}{p_{_F}\mp p_{_\alpha}}\Big),\nonumber
\end{eqnarray}
with $ W_0^{\pm} = \sqrt{(C_1^{\pm})^2 + (C_2^{\pm})^2}$.
Also, $\delta_0^{\pm} = \tan^{-1}(C_2^{\pm}/C_1^{\pm}) + \pi $.
The amplitudes, frequencies and phase factors are different 
for different branches. Also, the amplitudes, frequencies and phase factors 
depend on $n_e$, $\alpha $ and $a$. 
The diffusive conductivities for spin-up and spin-down electrons are
oscillating with different frequencies $\Omega^{\pm}$ in units of $\lambda$.
The total Weiss contribution to the  diffusive conductivity 
is given by 
\begin{eqnarray} \label{total_weiss}
\sigma_{\rm Weiss}^{\rm dif} & \backsimeq & \frac{e^2}{h} 
A_{_B} \lambda  
\Big[ C+  H\Big(\frac{T}{T_a}\Big) \Big\{ W_0^{+} 
\sin \big( 2\pi\Omega^{+} \lambda  
- \Phi_0^{+} \big) \nonumber \\ 
&+& W_0^{-}\sin \big( 2\pi\Omega^{-} \lambda  
- \Phi_0^{-} \big)\Big\}\Big],
\end{eqnarray}
where $C=C_0^{+}+ C_0^{-}$. 
It shows that the total diffusive conductivity exhibits
beating patterns in the amplitude of the Weiss oscillations.
In Fig. 2, we compare the analytical result of the Weiss 
contribution (\ref{total_weiss}) to the total 
diffusive conductivity with the exact numerical results obtained 
from Eq. (\ref{conduc}).
The analytical result are in excellent agreement with the
numerical result except at higher magnetic
field ($ \lambda \leq 0.15 $) where the SdH oscillation dominates over 
the magnetic Weiss oscillation.

To obtain the analytical expressions of the locations of the 
beat nodes and the number of oscillations between any two successive 
nodes, we simplify Eq. (\ref{total_weiss}) by considering 
$ p_{_F} \gg p_{\alpha}$ and $ 1/(p_{_F} \mp p_{\alpha}) \simeq 0 $.
Equation (\ref{total_weiss}) reduces to the following form:
\begin{eqnarray}\label{final}
\sigma_{\rm Weiss}^{\rm dif} & \simeq & \frac{e^2}{h}
\frac{ p_{_F}}{2 \pi^2}A_{_B}\lambda \Big[ 1 + H\Big(\frac{T}{T_a}\Big)  
\sin \Big( 2\pi\Omega_a \lambda + \pi \big) \nonumber \\
& \times &\cos \big( 2\pi\Omega_d \lambda \Big)\Big],
\end{eqnarray}
where $ \Omega_a = (\Omega_+ + \Omega_-)/2 $ and
$ \Omega_d = (\Omega_+ - \Omega_-)/2 $.

At the node positions $ B_0 = B_j $, we have the following
condition: $ \cos(2\pi \lambda \Omega_d)|_{B_0 = B_{j}} = 0 $, 
which gives us 
\begin{equation}\label{beat_con} 
B_{j} \backsimeq \frac{8 p_{_\alpha} B_{a}}{(2j+1)},
\end{equation}
where $j=0,1,2,3,4..$ are the $j$-th beat node and
the corresponding magnetic field is $B_j$. 
By using the above equation, the beating nodes appear at 
$B_a/B_j=0.11, 0.35, 0.59, 0.83, 1.07...$ which are in excellent
agreement with the exact numerical results. The above 
Eq. indicates that the magnetic field corresponding to the 
$j=0$ node is the lower limit of the Weiss 
oscillation below which the SdH oscillation starts to dominate. 
In practical, the numbering of the beat nodes is difficult. 
To remove this problem, the above equation can be re-written 
for any two successive beat nodes as
\begin{equation}\label{beat_loc}
\frac{1}{p_{_\alpha}}= 4B_a
\Big(\frac{1}{B_{j+1}}-\frac{1}{B_{j}}\Big).
\end{equation}
So, the strength of the Rashba SOI can be determined from the 
above equation by knowing the locations of the two successive beating nodes. 
The number of oscillation between any two successive beat nodes can be 
obtained from the first sine term of Eq. (\ref{final}), which is
\begin{equation} \label{num_osc}
N_{\rm osc} = 2p_{_F} B_{a} \Big(\frac{1}{B_{j+1}} 
- \frac{1}{B_{j}} \Big).
\end{equation}
Using Eq.(\ref{beat_loc}) in the above equation, we get
\begin{equation} \label{num_osc1}
N_{\rm osc}=\frac{p_{_F}}{2p_{_\alpha}} = \frac{k_{_F}}{2 k_{_\alpha}}.
\end{equation}
From the above equation, the following important conclusions can be drawn.
i) The above equation can be re-written as 
$ \alpha =  \hbar^2 k_{_F}/(2m^*N_{\rm osc}) $. 
Therefore, the Rashba SOI strength can be easily calculated by just counting 
the number of oscillation between any two successive nodes.
ii) The number of oscillation between any two successive nodes is constant
for given values of $n_e$ and $\alpha$, 
whereas it depends on the magnetic field in the SdH oscillation \cite{firoz}.
iii) While the frequency of the Weiss oscillation depends on the 
modulation period but the number of oscillation between any two successive 
nodes does not depend on it.

\begin{figure}[t]
\begin{center}\leavevmode
\includegraphics[width=97mm]{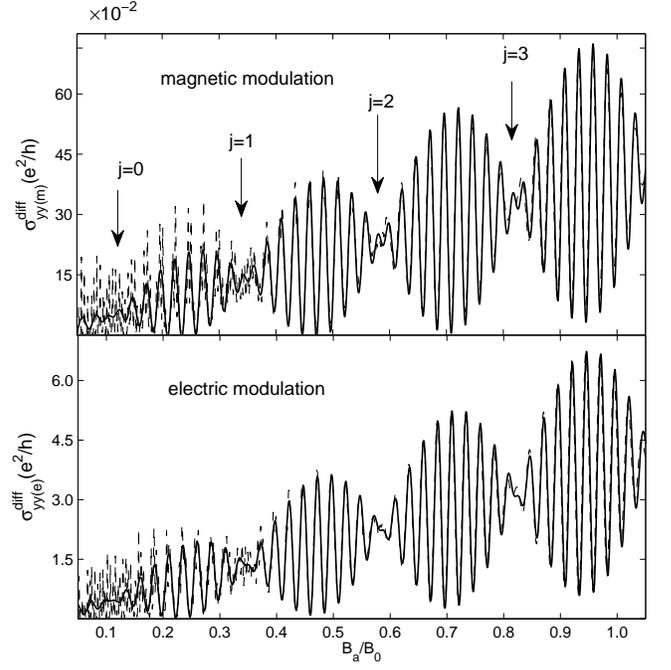}
\caption{Plots of the exact (dashed) and asymptotic (solid) expressions of 
the diffusive conductivity versus dimensionless inverse magnetic field
$B_a/B_0$ for electric and magnetic modulations.
 Here strength of the Rashba SOI  $\alpha=2\alpha_0$.}
\label{Fig2}
\end{center}
\end{figure}

The analytical expression of the diffusive conductivity given
in Eq. (\ref{total_weiss}) is not able to explain the origin 
of the superposition of the SdH oscillation on the Weiss oscillation 
at higher magnetic field. This can be explained by using the asymptotic
expression of the DOS given in Eq. (\ref{dos}) and replacing 
the summation \cite{vasilo} over discrete states by integration as 
$\sum_{s} \rightarrow 2\pi l_0^2\int_0^{\infty}D^{\pm}(E)dE$, then we get
$ \sigma_{yy}^{{\rm dif}(\pm)} = \sigma_{\rm Weiss}^{\pm} + \sigma_{\rm SdH}^{\pm} $,
where  $\sigma_{\rm Weiss}^{\pm} $ given 
in Eq. (\ref{diff_con}) and the Weiss oscillation
modulated by the SdH oscillation ($\sigma_{\rm SdH}^{\pm}$) is given by 
\begin{eqnarray} \label{diff_sdh}
\sigma_{\rm SdH}^{\pm} & = & \frac{e^2}{h} A_{_B} \lambda C_0^{\pm}
\exp{\Big\{-2\Big(\frac{\pi\Gamma_0}{\hbar\omega_0}\Big)^2\Big\}}
H\Big(\frac{T}{T_c}\Big) \nonumber \\ & \times & 
\Big[1+\Big(\frac{W^{\pm}}{C_0^{\pm}}\Big)
\sin \big(2\pi\Omega^{\pm} \lambda - \Phi^{\pm} \big)\Big] 
\nonumber \\ & \times & 
\cos(2\pi f^{\pm}/B_0),
\end{eqnarray}
where $ W^{\pm} = \sqrt{(C_1^{\pm})^2 + (2C_2^{\pm})^2}$,
$\Phi^{\pm}=\delta^{\pm} + 4\pi/(p_{_F}\mp p_{\alpha})$
with $\delta^{\pm} = \tan^{-1}(2C_2^{\pm}/C_1^{\pm}) + \pi $.
Also, $ f^{\pm} = (m^*/(\hbar e)) \big[ E_{F} + E_{\alpha}/2
\mp \sqrt{E_0^2+E_{\alpha} E_{F}} \big] $ are the
frequencies of the SdH oscillations for spin-up and spin-down
electrons in absence of the modulation \cite{firoz}
and the characteristic temperature for the SdH oscillation is 
$T_c=\hbar\omega_0/(2\pi^2k_{_B})$.
For $ \alpha = 2 \alpha_0 $ and $ T = 1.5 $ K, 
the ratio 
$ |\sigma_{{\rm SdH}}^{{\rm dif}}/\sigma_{{\rm Weiss}}^{{\rm dif}}| $ is 
0.3, 0.025, 0.001 for $ B_a/B_0 = 0.3, 0.5, 0.8$, respectively.   
Here, $\sigma_{{\rm SdH}}^{{\rm dif}} = \sigma_{{\rm SdH}}^{+}
+ \sigma_{{\rm SdH}}^{-}$.
Note that the Weiss oscillation frequencies ($\Omega^{\pm}$) for
spin-up and spin-down electrons are different from the SdH
oscillation frequencies $f^{\pm}$.

To derive Eqs. (\ref{diff_con}) and (\ref{diff_sdh}),
we have used the DOS for unmodulated 2DEG.
In the presence of the modulation, the DOS can be written as
\cite{con2002} a sum of the unmodulated ($D^{\pm}(E)$)  and 
modulated ($D_m^{\pm}(E)$) part as
$D_{\rm total}^{\pm}(E)= D^{\pm}(E) + D_m^{\pm}(E)$.
Here, $ D^{\pm}(E) $ is given in Eq. (\ref{dos}) and 
$ D_m^{\pm}(E) $ is of the order of $V_B^2$, which is very small 
in comparison to $D^{\pm}(E)$.
Moreover, the correction to the Weiss and the SdH oscillations
due to the $D_m(E)$ will be of the order of $ V_B^4$. 
The effect of the DOS correction due to the modulation on the Weiss and 
the SdH oscillations is really small and we have neglected it.

Substituting  $ \alpha =0 $ in Eq. (\ref{diff_con}), 
we get the oscillation period $ P = 1/\Omega = 1/(2 p_{_F})$
which is same as obtained in Ref. \cite{peeters}.
We compare this theoretical result with the available experimental result 
\cite{mag_exp3} for $\alpha = 0$.
In this experiment \cite{mag_exp3}, magnetic modulation
period $ a = 1 $ $\mu$m and
the electron density $ n_e = 2.2 \times 10^{15} $ /m${}^2$.
Using these parameters, we obtain $ P = 0.00425 $ which  
is very close to the experimental result $ P \simeq 0.00465$. 

{\bf Comparison with the electrical modulation case}: 
Here, we would like to compare the above mentioned results 
for the magnetically modulated case with the electrically 
modulated system. The electrical modulation potential
is described by $ H^{\prime} = V_E \cos(qx) $, where 
$V_E $ is the amplitude of the electric modulation potential.
We have used $ V_E = 0.05 $ meV in our numerical calculation.
In the case of electrically modulated system, numerical 
results have been discussed in Ref.\cite{wang}. Here, we 
provide an analytical expression of the diffusive conductivity 
for the electric modulation case.
We obtain the Weiss and the SdH contributions to the diffusive
conductivity, which are given below:
\begin{eqnarray} \label{diff_e}
\sigma_{\rm Weiss}^{\pm} & = & \frac{e^2}{h}
 \frac{A_{_E}\lambda}{(p_{_F}\mp p_{_\alpha})}
\cos^2 \{\pi/(p_{_F} \mp p_{_\alpha})\}
\Big[1 +  H\Big(\frac{T}{T_a}\Big)  \nonumber \\&\times&
\sin \Big\{ 2\pi\Omega^{\pm} \lambda  - 2\pi/(p_{_F}\mp p_{_\alpha})
 \Big\} \Big] 
\end{eqnarray}
and 
\begin{eqnarray}
\sigma_{\rm SdH}^{\pm} & = & \frac{e^2}{h}
\frac{A_{_E}\lambda}{(p_{_F}\mp p_{_\alpha})}
\cos^2 \{\pi/(p_{_F} \mp p_{\alpha})\} \nonumber \\ 
& \times &
\exp{\Big\{-2 \Big(\frac{\pi\Gamma_0}{\hbar\omega_0}\Big)^2\Big\}}
H\Big(\frac{T}{T_c}\Big) \nonumber \\ &\times &
\Big[1+\sin\{ 2\pi\Omega^{\pm} \lambda -
2\pi/(p_{_F}\mp p_{\alpha})\}\Big] \nonumber \\
& \times & \cos{(2\pi f^{\pm}/B_0)},
\end{eqnarray}
where $A_{_E}=V_{_E}^2\tau/(\hbar\eps_a)$. 
The lower panel of Fig. 2 shows the comparison between the
analytical expression given in Eq. (\ref{diff_e}) and the numerical results
reproduced from Eq. (20) of Ref. \cite{wang}. Our analytical
expression matches very well with the numerical result.
The Weiss and the SdH oscillation frequencies for spin-up and spin-down electrons
are the same for both type of the modulations. 
Therefore, Eqs. (\ref{beat_loc}), (\ref{num_osc}) and
(\ref{num_osc1}) for the magnetic modulation case remain also valid in the 
electric modulation case.

However, there are few important differences between the 
diffusive conductivities in the electric and magnetic modulation 
cases. There is a definite phase difference between the diffusive 
conductivities in electric and magnetic modulation cases.
The amplitude of the diffusive conductivity in the presence of 
the magnetic modulation is found to be nearly $[p_{_F}/(2\pi)]^2$ 
times higher in comparison to the electrical modulation case,
which is the same as in absence of the Rashba SOI \cite{peeters}.
The role played by the Rashba SOI in the case of magnetic modulation is that,
there is a difference between the amplitudes of conductivity for the
spin-up and spin-down electrons by a factor $p_{_\alpha}/(2\pi^2) $.
In the electric modulation case, amplitudes of the conductivities of spin up
and spin down branches are nearly the same.

\subsection{Collisional Conductivity}
In electron systems the charge impurities play an important 
role in magnetotransport properties. Collisional conductivity 
arises because of the migration of the cyclotron orbit due 
to scattering from charge impurities. At low temperature, we 
can assume that electrons are  elastically scattered  by the charged 
impurities distributed uniformly. The standard expression for 
collisional conductivity is given by\cite{vilet}
\begin{equation}
\sigma_{\mu \mu}^{\rm col} = \frac{\beta e^2}{2S_0}
\sum_{\xi, \xi^{\prime}} f_{\xi} (1-f_{\xi^{\prime}}) 
W_{\xi, \xi^{\prime}} 
(\alpha_\mu^{\xi} - \alpha_{\mu}^{\xi^{\prime}})^2.
\end{equation}
Here, $f_{\xi}=f_{\xi^{\prime}}$ for elastic scattering, 
$ W_{\xi,\xi^{\prime}} $ is the transition probability between 
one-electron states $ |\xi \ra $ and $ | \xi^{\prime} \ra $. 
Also, $ \alpha_{\mu}^{\xi} = \la \xi | r_{\mu} | \xi \ra $ is
the expectation value of the $\mu$ component of the position 
operator for the electron in state $ |\xi \ra $.
The scattering rate $ W_{\xi,\xi^{\prime}} $ is given by
\begin{equation}
W_{\xi,\xi^{\prime}} = \sum_{{\bf q}_{_0}} | U ({\bf q}_{_0})|^2 |\la \xi | 
e^{i {\bf q}_{_0} \cdot ({\bf r } - {\bf R})}| \xi^{\prime} \ra|^2 
\delta(E_{\xi} - E_{\xi^{\prime}}),    
\end{equation}
where $ {\bf q}_{_0}=q_{_{0x}} \hat x + q_{_{0y}} \hat y$ is 
a 2D wave-vector and 
$ U({\bf q}_{_0}) = 2 \pi e^2/
(\epsilon \sqrt{q_{_{0x}}^2 + q_{_{0y}}^2 + k_s^2}) $ 
is the Fourier transform of the screened impurity potential 
$ U({\bf r}) = (e^2/4\pi \epsilon) (e^{-k_sr}/r) $, where
$k_s$ is the inverse screening length and $ \epsilon $ is 
the dielectric constant of the material. 
In the limit of small $|{\bf q}_{_0}| \ll k_s $, 
$ U({\bf q}_{_0}) \simeq 2\pi e^2/(\epsilon k_s) = U_0$.
Here, ${\bf r} $ and ${\bf R} $ are the position vector of 
electron and impurity, respectively. In this limit, we can use
$\tau^2 \approx \pi l_0^2\hbar^2/N_IU_0^2$, 
where $N_I$ is the 2D impurity number density. 

We follow Refs. \cite{vasilo,peeters} to calculate the 
collisional conductivity.
We include the correction to the unperturbed eigenstate 
$|\Psi^{\pm}_{s,k_y}(r)\ra$ due to the weak 
perturbative term $ \Delta H=H_1+H_2+H_3+H_3$.
The first-order correction to the Landau state is obtained by
\begin{eqnarray}
|\Psi^{\pm}_{s,k_y}(r)\ra^{\prime} & = & |\Psi^{\pm}_{s,k_y}(r)\ra 
\nonumber \\
& + & \sum_{s^{\prime} \neq s} \frac{\la \Psi^{\pm}_{s,k_y}(r)|\Delta H|
\Psi^{\pm}_{s^{\prime},k_y}(r)\ra}{E_s-E_{s^{\prime}}}
|\Psi^{\pm}_{s^{\prime},k_y}(r)\ra \nonumber.
\end{eqnarray}
Following Refs. \cite{vasilo,peeters} and using the
the perturbed Landau states $|\Psi^{\pm}_{s,k_y}(r)\ra^{\prime} $, 
we obtain the collisional conductivity as
\begin{equation}\label{collisional}
\sigma_{xx}^{\pm} \approx \frac{e^2}{h}\frac{N_IU_0^2}{2\pi a\Gamma_0}
\sum_{s}[I_s^{\pm}M_s^{\pm}+R_s^{\pm}J_s^{\pm}].
\end{equation}
The exact expressions of 
$ I_s^{\pm}$ and $M_s^{\pm}$ are given as
\begin{eqnarray}
I_s^{\pm}&=&[(2s\mp1)D_s^4-2sD_s^2+2s\pm1]/A_s^2,\\
M_s^{\pm}&=&\int_{0}^{a/l_0^2}\Big[-\frac{\partial f(E)}{\partial E}
\Big]_{E=E^{\pm}_{s,k_y}}dk_y,\\
J_s^{\pm} & = &\int_{0}^{a/l_0^2}\Big[-\frac{\partial f(E)}{\partial E}
\Big]_{E=E^{\pm}_{s,k_y}}\sin^2(qx_0)dk_y.
\end{eqnarray}
The term $R_s^{\pm}$ is appearing due to the first-order correction
to the Landau wave function and it is of the order of $V_B^2$. 
We neglect this term because of small contribution.
The major contribution to the collisional conductivity is due
to the first term proportional to $M_s^{\pm}$. The effect
of the magnetic modulation mainly enters through the 
energy correction due to the modulation 
in the total energy in the Fermi-Dirac distribution function.

Similar to the diffusive conductivity, the collisional conductivity
in the presence of the modulation will have two contributions,
namely the SdH and the Weiss contributions:
$ \sigma_{xx}^{\pm} = \sigma_{\rm SdH}^{\pm} + \sigma_{ \rm Weiss}^{\pm}$.
It is difficult to get the analytical expression of the Weiss
contribution $(\sigma_{\rm Weiss}^{\pm})$ comes from the energy correction
in Fermi-Dirac distribution function.

The numerical results of the change in the collisional conductivity 
$ \Delta\sigma_{xx} $
versus magnetic field for  $\alpha = 2 \alpha_0$ are plotted in Fig. (3).
To compare the results of the magnetic modulation case, we 
present $\Delta\sigma_{xx} $ for the electric modulation case in the
lower panel of Fig. (3).

\begin{figure}[t]
\begin{center}\leavevmode
\includegraphics[width=95mm]{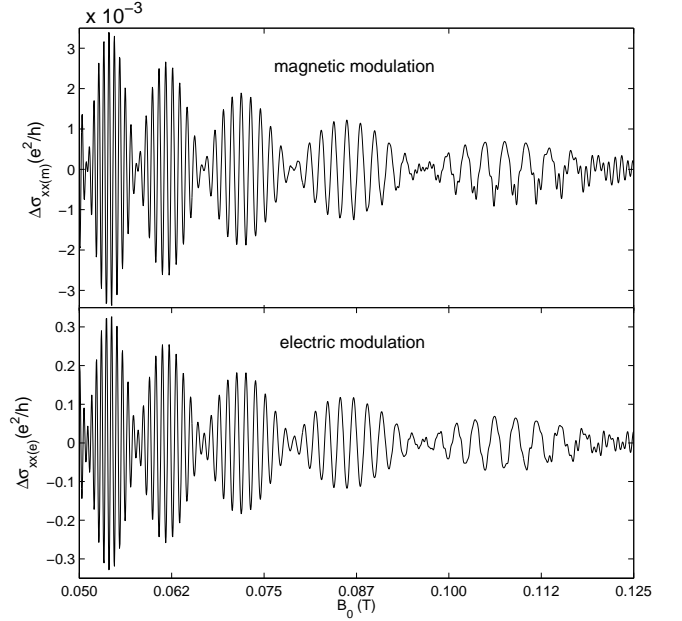}
\caption{Plots of the modulation induced change in the collisional 
conductivity versus magnetic field for $\alpha = 2 \alpha_0$.}
\label{Fig5}
\end{center}
\end{figure}
The oscillatory behavior with beating pattern appears in the 
changes in the conductivity due to the modulation at low magnetic field
range where modulation strength is not much
less than the energy scale of the Landau levels. The effect of
the modulation diminishes with the increase of the strength of the 
perpendicular magnetic field. As the magnetic field increases, the SdH 
oscillation starts to dominate over the modulation induced Weiss oscillation. 

The collisional conductivities for spin-up and spin-down electrons
oscillate with the same frequencies 
as the bandwidth $ |\Delta_s^{\pm}|$. The frequencies of the
oscillation of the bandwidths are the same as that of the diffusive
conductivities, namely, $\Omega^{\pm}$.
Therefore, beating condition for the collisional conductivity is 
the same as given in Eq. (\ref{beat_con}) for the diffusive conductivity.
Using the beating condition, we get the magnetic fields (in Tesla) 
corresponding to the beating positions, obtained from the Eq. (\ref{beat_con}), 
are 0.0506, 0.0574, 0.0662, 0.0782, 0.0957....
These are in good agreement with the exact numerical results shown in Fig. 3.
The phase difference between the oscillations in
the diffusive and collisonal conductivities is nearly $\pi$.

The modulation gives very small effect on the SdH part
$ (\sigma_{\rm SdH}^{\pm}) $ of the collisional conductivity.
It is also difficult to get an analytical expression of 
the SdH oscillations superposed on the Weiss oscillations. 
Ignoring the modulation effect and using the analytic form
of the DOS, the asymptotic expression of the SdH oscillation 
has been studied in Ref. \cite{firoz}.

The beating pattern in the SdH oscillation is given by \cite{firoz}
\begin{eqnarray} 
\frac{\sigma_{_{SdH}}}{\sigma_0} &  \simeq & 
\frac{\tilde{E}_{F}}{4(\omega_0 \tau)^2} \Big[1 + 
2\exp{\Big\{-2\Big(\frac{\pi\Gamma_0}
{\hbar\omega_0}\Big)^2\Big\}}  H\Big(\frac{T}{T_c}\Big) \nonumber \\
& \times & \cos(2\pi f_{a}/B_0) \cos(2\pi f_{d}/B_0)\Big],
\end{eqnarray}
where $\sigma_0=n_e e^2\tau/m^*$ is the classical Drude conductivity,
$\tilde{E}_{F} = [1 + E_{\alpha}/(2E_{F}) \mp
(3/2)\sqrt{E_{\alpha}/E_{F}}]$ and
$ f_{a} = (f^+ + f^-)/2 $ and $ f_{d} = (f^+ - f^-)/2 $.

\subsection{The Hall conductivity} 
The off-diagonal elements in the conductivity 
tensor are termed as the Hall conductivity which is given by\cite{vilet}
\begin{eqnarray}\label{hall} 
\sigma_{yx} &=& \frac{i e^2 \hbar}{S_0} \sum_\xi f_{\xi}(1-f_{\xi^{\prime}}) 
\la \xi| v_y|\xi^{\prime}\ra \la \xi^{\prime}|v_x|\xi\ra\nonumber\\
&\times&\frac{1-e^{\beta(E_{\xi}-E_{\xi'})}}{{(E_{\xi} - E_{\xi^{\prime}})^2}}.
\end{eqnarray}
To simplify the above equation, we shall use the following relation
\begin{equation}
f_{\xi}(1-f_{\xi^{\prime}})
\Big[1-e^{\beta(E_{\xi}-E_{\xi^{\prime}})}\Big]=f_{\xi}-f_{\xi^{\prime}}.
\end{equation}

The matrix elements of the velocity operators are zero except 
between the nearest Landau levels. Following Refs. 
\cite{vasilo,peeters,vasi}, 
we calculate the velocity matrix 
elements and the energy difference between two nearest Landau levels
(see Appendix B). 
Finally, we get the expressions of the Hall conductivity 
as
\begin{eqnarray} \label{hall}
\sigma_{yx}^{\pm}&=&\frac{e^2}{h}\frac{l_0^2}
{a}\sum_s\frac{1}{{A_s A_{s+1}}}\Big[D_{s+1} \big(D_s \sqrt{s} 
\pm \sqrt{2} k_{\alpha} l_0 \big) \nonumber \\ 
& + & \sqrt{s+1} \Big]^2 \int_0^{a/{l_0}^2}
\frac{f_{s,k_y}-f_{s+1,k_y}}
{[1 + \Upsilon_s^{\pm}\cos{qx_0}]^2}dk_y \label{hallcon}.
\end{eqnarray}

In the large $s$ limit, $\Upsilon_s^{\pm} $ reduces to 
\begin{eqnarray} \label{upsilon}
\Upsilon_s^{\pm} & \simeq & \frac{V_{_B}\sqrt{\lambda}}{\pi\epsilon_a}
\frac{(p_{_F} \mp p_{\alpha})^{3/2}}{\pi^2}\sin^2\{\pi/(p_{_F}\mp p_{_\alpha})\} 
\nonumber\\
& \times & \cos\Big\{ 2 \pi \lambda (p_{_F} \mp p_{\alpha})
- \frac{\pi}{4} -\frac{\pi}{(p_{_F}\mp p_{_\alpha})} \Big\}.
\end{eqnarray}

\begin{figure}[t]
\begin{center}
\includegraphics[width=96mm]{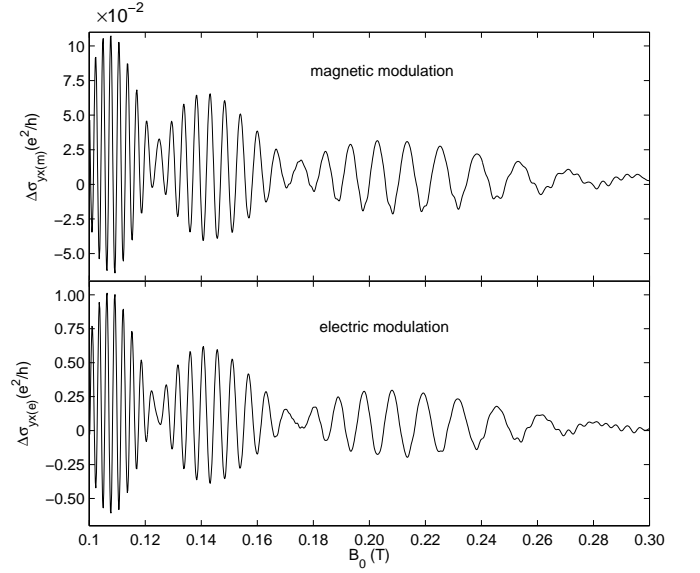}
\caption{Plots of the change in the Hall conductivity due to modulation 
versus magnetic field $B_0$ for  $\alpha=2\alpha_0 $.}
\label{Fig6}
\end{center}
\end{figure}

When $ T \rightarrow 0 $ and $ E_s < E_F < E_{s+1}$,
Eqs. (\ref{hallcon}) and (\ref{upsilon}) give the 
asymptotic form of the Hall conductivity as
\begin{eqnarray}
\sigma_{yx}^{\pm}  \simeq \frac{\sigma_{_{\rm H,\alpha}}}{2}  
[1 + \frac{3}{2} (\Upsilon_{s^{^\pm}}^{\pm})^2],
\end{eqnarray}
where 
\begin{equation}
\sigma_{_{\rm H,\alpha}} =
\sigma_{_{\rm H}}\Big[1-\frac{1}{2}\Big(\frac{k_{\alpha}}{k_F}\Big)^2\Big]^2 
\backsimeq
\sigma_{_{\rm H}}\Big[1-\Big(\frac{k_{\alpha}}{k_F}\Big)^2\Big] 
\end{equation}
is the total Hall conductivity of the 2DEG with Rashba SOI but without modulation 
and  $\sigma_{_{\rm H}} = n_e e/B_0 $ is the classical Hall conductivity. Note that 
$n_e$ is the sum of the density of the spin-up and spin-down electrons.  
By taking the superposition of the analytical expression of the
conductivity for the spin-up and spin-down electrons, 
we get the same beating condition as given in Eq. (\ref{beat_con}) for
the diffusive conductivity. The numerical results of the change in the Hall conductivity
versus magnetic field is shown in figure 4.

\begin{figure}[t]
\begin{center}
\includegraphics[width=96mm]{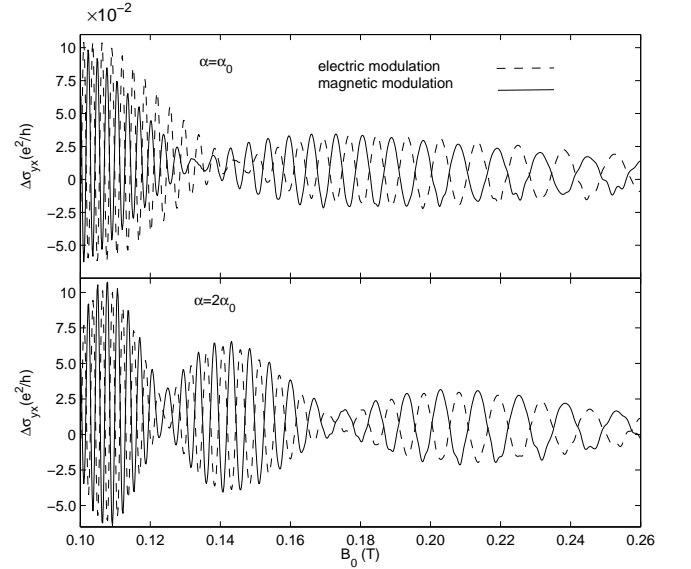}
\caption{Plots of the change of the Hall conductivity 
versus magnetic field $B_0$ for different values of $\alpha$. 
We have scaled up the electrical modulation case by 10.}
\label{Fig6}
\end{center}
\end{figure}

Similarly, we obtain the Hall conductivity for the electrical modulation case as
$ \sigma_{yx}^{\pm}  \simeq \frac{\sigma_{_{\rm H,\alpha}}}{2}
[1 +  (3/2)(\Upsilon_{s,e}^{\pm})^2]$, where
\begin{eqnarray}
\Upsilon_{s(e)}^{\pm} & \simeq & \frac{V_E}{\pi\epsilon_a }
\sqrt{\frac{\lambda}{p_{_F} \mp p_{\alpha}}}
\sin\{2\pi/(p_{_F}\mp p_{_\alpha})\} \nonumber \\
& \times & \sin\Big\{ 2 \pi \lambda (p_{_F} \mp p_{\alpha})-\pi/4\Big\}.
\end{eqnarray}
In figure 5, we compare the change in the Hall conductivity of
 the electric and magnetic modulation case.
The beating condition remains the same in the electric modulation case also.
The amplitude in the change in conductivity due to the magnetic 
modulation is nearly  $[p_{_F}/(2\pi)]^2$
times higher than the electrical modulation case.
There is a $\pi$ phase difference between the Hall conductivities of
the electrical and magnetic modulation cases. In the above
analytical expression modulation effect through the Fermi-Dirac 
distribution function has been ignored.

\section{CONCLUSION}
We have studied magnetotransport properties
of the 2DEG in the presence of the Rashba SOI
when the perpendicular magnetic field is weakly modulated. 

The diffusive conductivity shows beating pattern due to the interference 
between the conductivities for spin-up and spin-down electrons.
We calculate the asymptotic expression of the Weiss conductivity 
which matches very well with the exact numerical results.
The number of Weiss oscillations between any two successive nodes
is fixed for given values of $n_e$ and $\alpha$ where as in the 
SdH it depends on the magnetic field.
We have shown that the Rashba SOI strength can be determined by 
analyzing the beating pattern in the Weiss oscillation.
The strength of the Rashba SOI can be determined by just 
counting the number of Weiss oscillations between any two 
successive beat nodes.
There is a definite phase difference between the conductivities for
magnetic and electric modulation cases. In magnetically modulated system,
there is a difference in amplitudes by a factor $p_{_\alpha}/(2\pi^2)$ 
between the conductivity due to the spin-up and spin-down electrons 
where as in the electrically modulated system, the amplitudes are the same 
for spin-up and spin-down electrons.

To observe the effect of the modulation,  we plot the
change in the collisional conductivity due to the modulation at
the low range of the magnetic field.  It is found that the effect of 
the magnetic modulation is much higher than the electrical modulation.
The major effect of the modulation comes through the energy correction in
the total energy in the Fermi-Dirac distribution function.
The beating condition in the Weiss contribution to the collisional
conductivity is same as that of the diffusive conductivity.

The  modulation effect on the Hall conductivity is shown by plotting 
the change in the Hall conductivity due to the modulation.
The beating pattern appears in the Hall conductivity and it increases with
the increase of the Rashba strength. The oscillations are out of phase
between the electric and magnetic modulation cases. The beating 
condition remains same as the diffusive conductivity.
The amplitude of the fluctuation in the presence of the magnetic modulation is
found to be much higher in comparison to the electric modulation case.

\section{acknowledgement}
This work is financially supported by the CSIR, Govt. of India under the grant
CSIR-SRF-09/092(0687) 2009/EMR F-O746.

\begin{appendix}

\section{}
We derive the DOS by taking the imaginary part of 
the self-energy \cite{ando,gerhartz} which is given as
\begin{equation}
\Sigma^-(E)=\Gamma_0^2\sum_s \frac{1}{E-E_{s}-\Sigma^-(E)}.
\end{equation}
The DOS is the imaginary part of the self-energy:
$ D(E)= Im \left[\frac{\Sigma^-(E)}{\pi^2 l_0^2 \Gamma_0^2}\right] $.
First we consider the lower branch.
To find the summation, we use the residue theorem and neglecting
$ (E_{\alpha}/\hbar \omega_0)^2 $ term,
we obtain
$ \Sigma^-(E)=\frac{\pi\Gamma_0^2}{\hbar\omega_0}\cot(\pi s_+) $, 
where
$s_+ \simeq \frac{1}{\hbar\omega_0}\{E-\Sigma^{-}(E)+
E_{\alpha}/2+\sqrt{E_0^2+E_{\alpha}E}\}$.
We write $ \Sigma^-(E) = \Delta + i \Gamma/2 $, 
then the above equation can be re-written as
\begin{equation}
\Delta + i \Gamma/2 = \frac{\pi\Gamma_0^2}{\hbar\omega_0} 
\cot\Big[\frac{(u-iv)}{2} \Big]
= \Big(\frac{\pi\Gamma_0^2}{\hbar\omega_0} \Big) 
\frac{\sin u + i \sinh v}{\cosh v - \cos u}.\\
\end{equation}
Here,
$ u = \frac{2\pi}{(\hbar\omega_0)}\{E-\Delta+E_{\alpha}/2 + 
\sqrt{E_0^2+E_{\alpha}E}\} $ and
$ v = \pi\Gamma/(\hbar\omega_0) $.
The imaginary part is
$ \frac{\Gamma}{2} = \Big(\frac{\pi\Gamma_0^2}{\hbar\omega_0}\Big) 
\frac{\sinh v}{\cosh v-\cos u} $.
We are using the following standard result to simplify it further:
\begin{equation}
\frac{\sinh v}{\cosh v-\cos u} = 1 + 2 \sum_{k=1}^{\infty} e^{-k v} \cos (ku).
\end{equation}
We are considering the first term corresponding to $k=1 $ only. 
Other terms are very small compared to the first one. We have
$
\frac{\Gamma}{2}  = \Big( \frac{\pi\Gamma_0^2}{\hbar\omega_0}\Big)
\Big [1 + 2  e^{-\pi \Gamma/(\hbar \omega_0)} \cos(u) \Big].
$
In the limit of $\pi\Gamma \gg \hbar\omega_0 $, after first iteration, we have
$ \Gamma/2=\pi\Gamma_0^2/(\hbar\omega_0) $.
We are putting it back in the earlier expression, we get
\begin{eqnarray}
\frac{\Gamma}{2} & = & \Big(\frac{\pi\Gamma_0^2}{\hbar\omega_0}\Big)
\Big [1+2 \exp\Big\{-2\Big(\frac{\pi\Gamma_0}{\hbar\omega_0}\Big)^2\Big\} 
\nonumber \\
& & 
\cos\Big\{\frac{2\pi}{\hbar\omega_0}
(E+E_{\alpha}/2+\sqrt{E_0^2+E_{\alpha}E}) \Big\} \Big].
\end{eqnarray}
Similarly, one can do it for the upper branch.
Finally, the DOS for lower and upper branches can be put together as
\begin{eqnarray}
D^{\pm}(E)&=&\frac{m^*}{2\pi \hbar^2}\Big[1+2\exp\Big\{-2\Big(\frac{\pi\Gamma_0}
{\hbar\omega_0}\Big)^2\Big\}\nonumber\\&\times&
\cos \Big\{\frac{2\pi}{\hbar\omega_0}
\Big(E+\frac{E_{\alpha}}{2}\mp\sqrt{E_0^2+E_{\alpha}E}\Big)\Big\}\Big].
\end{eqnarray}
In absence of the Rashba SOI and the Zeeman term, including the spin degeneracy the
above expression can be reduced to the following standard result 
\cite{ando,gerhartz} as
\begin{equation}
D(E)=\frac{m^*}{\pi \hbar^2}\Big[1+2\exp\Big\{-2\Big(\frac{\pi\Gamma_0}
{\hbar\omega_0}\Big)^2\Big\}\cos \Big(\frac{2\pi E}{\hbar\omega_0}-\pi\Big)\Big].
\end{equation}

\section{}
The velocity operators are given by
\begin{equation}
v_x=\frac{\partial H_0}{\partial p_{x}}=
\frac{p_x}{m^*}-\frac{\alpha}{\hbar}\sigma_y=
\begin{bmatrix}{}\frac{p_{x}}{m^*} & i\frac{\alpha}{\hbar}\\ 
-i\frac{\alpha}{\hbar} & \frac{p_{x}}{m^*}
\end{bmatrix}.
\end{equation}

\begin{eqnarray}
v_y&=&\frac{\partial H_0}{\partial p_{y}}= 
\frac{1}{m^*}(p_{y}+eB_0x)+\frac{\alpha}{\hbar}\sigma_x
\nonumber\\
&=&\begin{bmatrix}{}\frac{p_y+eB_0x}{m^*} & \frac{\alpha}{\hbar}\\ 
\frac{\alpha}{\hbar} & \frac{p_{y}+eB_0x}{m^*}
\end{bmatrix}.
\end{eqnarray}

For $\xi^{\prime}: \{s+1,\pm,k_y\}$, 
the velocity matrix elements (see Refs. \cite{vasilo,peeters,vasi}) are
\begin{eqnarray}\label{vx}
\la \Psi^{\pm}_{s,k_y}(r)\mid &v_x&\mid \Psi^{\pm}_{s+1,k_y}(r)\ra =
\frac{1}{\sqrt{A_s A_{s+1}}}\Big[D_s D_{s+1}\nonumber \\ 
& \times &
\la \phi^{\pm}_{s-1}(X) \mid \frac{p_{_x}}{m^*} \mid \phi^{\pm}_{s}(X)\ra 
\nonumber \\ 
& + &
\la \phi^{\pm}_{s}(X) \mid \frac{p_{_x}}{m^*} \mid \phi^{\pm}_{s+1}(X)\ra\
\mp i\frac{\alpha}{\hbar}D_{s+1}\Big]\nonumber\\
&=& \frac{-i}{\sqrt{A_sA_{s+1}}}\sqrt{\frac{\hbar\omega_0}{2m^*}}
\Big[D_{s+1}(D_s\sqrt{s}\pm\sqrt{2}k_{\alpha}l_0)
\nonumber\\&+&\sqrt{s+1}\Big],
\end{eqnarray}
where $ X=x+x_0$.
Similarly, for the velocity component $v_y$ 
\begin{eqnarray}\label{vy}
\la \Psi^{\pm}_{s+1,k_y}(r)\mid &v_y&\mid \Psi^{\pm}_{s,k_y}(r)\ra=
\frac{1}{\sqrt{A_sA_{s+1}}}\sqrt{\frac{\hbar\omega_0}{2m^*}}\nonumber\\&\times&
\Big[D_{s+1}(D_s\sqrt{s}\pm\sqrt{2}k_{\alpha}l_0)
+\sqrt{s+1}\Big].
\end{eqnarray}
The energy difference between the two successive Landau levels is
\begin{equation}\label{deltaE}
E_{s,k_y}-E_{s+1,k_y} \simeq - \hbar\omega_0 [1-\Upsilon_s^{\pm}\cos(qx_0)],
\end{equation}
where $\Upsilon_s^{\pm}=(F_{s+1}^{\pm}-F_{s}^{\pm})/(\hbar\omega_0)$. 
Substituting the above three equations (\ref{vx}), (\ref{vy}) and (\ref{deltaE}) 
in the Hall conductivity expression, we get Eq. (\ref{hall}).
A multiplication factor 2 need to be used for the contribution coming from 
$\xi^{\prime}: \{s-1,\pm,k_y\}$ states.

\end{appendix}

\end {document}